\documentclass[twocolumn,showpacs,fleqn,nobibnotes]{revtex4}

\usepackage{amsmath}
\usepackage{graphicx}
\usepackage{float}
\usepackage{subfigure}

\def\lsim{\raise0.3ex\hbox{$<$\kern-0.75em\raise-1.1ex\hbox{$\sim$}}}
\def\gsim{\raise0.3ex\hbox{$>$\kern-0.75em\raise-1.1ex\hbox{$\sim$}}}

\newcommand{\rr}{\mbox{\boldmath $r$}}

\newcommand{\bb}{\mbox{$b$}}

\begin{document}

\title{Diffractive photoproduction of heavy quarks in hadronic collisions}
\author{V.P. Gon\c{c}alves
$^{a}$
and M.V.T. Machado $^{b}$ }

\affiliation{$^a$ Instituto de F\'{\i}sica e Matem\'atica, Universidade Federal de
Pelotas\\
Caixa Postal 354, CEP 96010-090, Pelotas, RS, Brazil\\
$^b$ Centro de Ci\^encias Exatas e Tecnol\'ogicas, Universidade Federal de Pelotas \\
Campus de Bag\'e, Rua Carlos Barbosa. CEP 96400-970. Bag\'e, RS, Brazil }

\begin{abstract}
In this letter we study the diffractive photoproduction of heavy quarks in hadronic ($pp/pA/AA$)  interactions for  Tevatron and LHC energies. The integrated cross section and rapidity distribution for the process $h_1 h_2 \rightarrow h_1 h_2 Q \overline{Q}$ ($h_i = p,A$ and $Q = c,b$)  are estimated using the Color Glass Condensate (CGC) formalism.  Our results indicate that this production channel has larger cross sections than the competing reactions of  double diffractive  production and coherent $AA$ reactions initiated by two-photon collisions.

\end{abstract}

\maketitle

{\it Introduction}. 
The QCD dynamics at high energies is of utmost importance for building a realistic description of $pp/pA/AA$ collisions at LHC. Theoretically, at high energies the QCD evolution leads to a system with high gluon density, characterized by the limitation on the maximum phase-space parton density that can be reached in the hadron
wavefunction (parton saturation). The transition is specified by a typical scale, which is energy dependent and is called saturation scale $Q_{\mathrm{sat}}$ (For recent reviews see Ref. \cite{hdqcd}).
Signals of parton saturation have already
been observed both in  $ep$ deep inelastic scattering at HERA and in deuteron-gold 
collisions at RHIC (See, e.g. Refs. \cite{blaizot,vicmag_plb}). However, the observation of this new regime still 
needs confirmation and so there is an active search for new experimental signatures. 
Among them, the observables measured in diffractive processes   deserve special 
attention. As shown in Ref. \cite{GBW}, the total diffractive cross section is much more sensitive to 
large-size dipoles than the inclusive one. As saturation effects screen large-size dipole (soft) contributions, one has that a fairly large fraction of the cross section is hard and hence eligible for a perturbative treatment. Therefore,  the study of diffractive processes becomes fundamental in order to constrain the QCD dynamics at high energies.

In this paper we propose to study diffractive interactions in ultra-peripheral collisions of hadrons, which can be defined as collisions where no hadronic interactions occur because the large spatial separation between projectile and target and the interaction is mediated by the electromagnetic field (For recent reviews see Ref. \cite{upcs}). In particular, we analyze the {\it diffractive} heavy quark photoproduction in $pp/pA/AA$ collisions, which at the Large Hadron Collider (LHC) will allow photon-hadron interactions to be studied at energies higher than at any existing accelerator.
In relativistic heavy ion colliders, the heavy nuclei give rise to strong electromagnetic fields, which can interact with each other. In a similar way, these processes also occur when considering energetic protons in $pp(\bar{p})$ colliders. Over the past years a comprehensive analysis of the {\it inclusive} heavy quark \cite{antigos,vicber,klein_vogt,per2,per3,vicmag_prd,frank}  production in ultraperipheral heavy ion collisions was made considering different theoretical approaches. As a photon stemming from the electromagnetic field
of one of the two colliding nuclei can interact with one photon of
the other nucleus (two-photon process) or can penetrate into the
other nucleus and interact with its hadrons (photon-nucleus
process), both possibilities has been studied in the literature. In principle, the experimental signature of these two processes is distinct and it can easily be separated. While in two-photon interactions we expect the presence of two rapidities gaps and no hadron breakup, in the inclusive heavy quark photon-hadron production the hadron target we expect only one rapidity gap and the dissociation of the hadron.  One of the main motivations to analyze the diffractive heavy quark photoproduction is that we expect the presence of two rapidity gaps in the final state, similarly to two-photon interactions. Consequently, it is important to determine the magnitude of this cross section in order to estimate the background for two-photon interactions.
As discussed in Refs. \cite{klein_vogt,per2,per3}, the heavy quark production in  $\gamma \gamma$ interactions is approximately two or three orders of magnitude smaller than the inclusive photoproduction cross sections.  However, the magnitude of the diffractive photoproduction cross-section is still an open question. Another motivation for our study is that the contribution of this process can be important in proton-proton collisions, where there is a dedicated program to search evidence of the Higgs and/or new physics in central double diffractive production processes \cite{martin}, which also are characterized by two rapidity gaps and has as main background the exclusive $b\overline{b}$ production. 

{\it Ultra-peripheral collisions}. 
The basic idea in ultra-peripheral hadron collisions is that the total 
cross section for a given process can be factorized in terms of the equivalent flux of photons of the hadron projectile and  the photon-photon or photon-target production cross section \cite{upcs}. 
In particular, the photon-hadron interactions can be divided into exclusive and inclusive reactions. In the first case, a certain particle is produced while the target remains in the ground state (or is only internally excited). On the other hand, in inclusive interactions the particle produced is accompanied by one or more particles from the breakup of the target. The typical examples of these processes are the exclusive vector meson production, described by the process $\gamma h \rightarrow V h$ ($V = \rho, J/\Psi, \Upsilon$), and the inclusive heavy quark production [$\gamma h \rightarrow X Y$ ($X = c\overline{c}, b\overline{b}$)], respectively. In the last years we have discussed in detail both processes considering $pp$ \cite{per4,vicmag_prd}, $pA$ \cite{vicmag_prc} and $AA$ \cite{per3,per4} collisions as an alternative to constrain the QCD dynamics at high energies. Here we propose to analyze another exclusive process, characterized by the diffractive photoproduction of heavy quarks and described by the   $\gamma h \rightarrow X h$ reaction. In this case, the cross section for the diffractive photoproduction of a final state $X$ in a ultra-peripheral  hadron-hadron collision is given by
\begin{eqnarray}
\sigma (h_1 h_2 \rightarrow h_1 h_2 X)\, = \int \limits_{\omega_{min}}^{\infty} d\omega \frac{dN_{\gamma}(\omega)}{d\omega}\,\sigma_{\gamma h \rightarrow X h} \left(W_{\gamma h}^2\right)\,,
\label{sigAA}
\end{eqnarray}
where 
$\omega$ is the photon energy   and $ \frac{dN_{\gamma}(\omega)}{d\omega}$ is the equivalent flux of photons from a charged hadron. Moreover, $\omega_{min}=M_{X}^2/4\gamma_L m_p$, $\gamma_L$ is the Lorentz boost  of a single beam,  $W_{\gamma h}^2=2\,\omega\sqrt{S_{\mathrm{NN}}}$  and
$\sqrt{S_{\mathrm{NN}}}$ is  the c.m.s energy of the
hadron-hadron system. It is important to emphasize that the equivalent photon energies at the LHC will be higher than at any existing accelerator. For instance, considering $p Pb$ collisions at LHC, the Lorentz factor  is
$\gamma_L= 4690$, giving the maximum c.m.s. $\gamma h$ energy
$W_{\gamma p} \approx 1500$ GeV. Therefore, while studies of photoproduction at HERA are limited to photon-proton center of mass energies of about 200 GeV, photon-hadron interactions at  LHC can reach one order of magnitude higher on energy. Consequently, studies of $\gamma h$ interactions at LHC could provide valuable information on the QCD dynamics at high energies.  In this work we consider that the produced state $X$ represents a  $Q\overline{Q}$ pair. Since photon emission is coherent over the entire proton/nucleus and the photon is colorless we expect that the diffractive events to be characterized by two  rapidity gaps, in contrast with the inclusive heavy quark production. In  these two-rapidity gaps events the heavy quark pair is produced in the central rapidity region, whereas the beam particles often leave the interaction region intact, and can be measured using very forward detectors.

In the calculations what follows we consider that the photon spectrum for a nuclei is given by \cite{upcs}
\begin{eqnarray}
\frac{dN_{\gamma}\,(\omega)}{d\omega}= \frac{2\,Z^2\alpha_{em}}{\pi\,\omega}\, \left[\bar{\eta}\,K_0\,(\bar{\eta})\, K_1\,(\bar{\eta})+ \frac{\bar{\eta}^2}{2}\,{\cal{U}}(\bar{\eta}) \right]\,
\label{fluxint}
\end{eqnarray}
where $\bar{\eta}=\omega\,R_{eff}/\gamma_L$ and  ${\cal{U}}(\bar{\eta}) = K_1^2\,(\bar{\eta})-  K_0^2\,(\bar{\eta})$, with $R_{eff} = R_p + R_A$  ($R_{eff} = 2 R_A$) for $pA$ ($AA$) collisions. On the other hand, for a  proton, we assume that the  photon spectrum is given by  \cite{Dress},
\begin{eqnarray}
\frac{dN_{\gamma}(\omega)}{d\omega} =  \frac{\alpha_{\mathrm{em}}}{2 \pi\, \omega} \left[ 1 + \left(1 - 
\frac{2\,\omega}{\sqrt{S_{NN}}}\right)^2 \right] \nonumber \\
\left( \ln{\Omega} - \frac{11}{6} + \frac{3}{\Omega}  - \frac{3}{2 \,\Omega^2} + \frac{1}{3 \,\Omega^3} \right) \,,
\label{eq:photon_spectrum}
\end{eqnarray}
with the notation $\Omega = 1 + [\,(0.71 \,\mathrm{GeV}^2)/Q_{\mathrm{min}}^2\,]$ and $Q_{\mathrm{min}}^2= \omega^2/[\,\gamma_L^2 \,(1-2\,\omega /\sqrt{S_{NN}})\,] \approx (\omega/
\gamma_L)^2$.

{\it QCD dynamics at high energies}. 
The photon-hadron interaction at high energy (small $x$) is usually described in the infinite momentum frame  of the hadron in terms of the scattering of the photon off a sea quark, which is typically emitted  by the small-$x$ gluons in the proton. However, in order to describe diffractive interactions and disentangle the small-$x$ dynamics of the hadron wavefunction, it is more adequate to consider the photon-hadron scattering in the dipole frame, in which most of the energy is
carried by the hadron, while the  photon  has
just enough energy to dissociate into a quark-antiquark pair
before the scattering. In this representation the probing
projectile fluctuates into a
quark-antiquark pair (a dipole) with transverse separation
$\rr$ long after the interaction, which then scatters off the target \cite{nik}. The main motivation to use this color dipole approach is that it gives a simple unified picture of inclusive and diffractive processes. In particular,  in this approach  the diffractive heavy quark photoproduction cross section $[\gamma h \rightarrow Q\overline{Q}h, \,\,h = p,A]$ reads as,
\begin{eqnarray}
\sigma^D_{T,L}= 
\int d^2\bb \, dz\, d^2\rr \,|\Psi^{\gamma}_{T,L} (z,\rr,Q^2)|^2\,{\cal{N}}^2(\bar{x},\rr,\bb) \,\,,
\end{eqnarray} 
where 
$\Psi^{\gamma}_{T,L}$  is the light-cone wavefunction  of the photon \cite{nik}.  The variable $\rr$ defines the relative transverse
separation of the pair (dipole) and $z$ $(1-z)$ is the
longitudinal momentum fractions of the quark (antiquark). The basic
blocks are the photon wavefunction, $\Psi^{\gamma}$  and the dipole-target forward amplitude ${\cal{N}}$. For photoproduction we have that longitudinal piece does not contribute, since $|\Psi_{L}|^2\propto Q^2$, and the total cross section is computed introducing the appropriated mass and charge of the charm or  bottom quark.

\begin{table}[t]
\begin{center}
\begin{tabular} {||c|c|c|c||}
\hline
\hline
$h_1 h_2$ & Collider   & $c\overline{c}$ & $b\overline{b}$  \\
\hline
\hline
$pp(\overline{p})$  & RHIC &  3.4 nb  & 3 $\times 10^{-3}$ nb \\
\hline
  & TEVATRON &  12.6 nb  & 0.021 nb \\
\hline
  & LHC &  92.0 nb  & 0.2 nb \\
\hline
\hline
$pA$  & LHC &  54.0 $\mu$b  & 0.09 $\mu$b \\
\hline
\hline
$AA$  & LHC &  59.0 mb  & 0.01 mb \\
\hline
\hline
\end{tabular}
\end{center}
\caption{\it The integrated cross section for the diffractive photoproduction of heavy quarks in $pp/pA/AA$ collisions.}
\label{tabhq}
\end{table}


In the Color Glass Condensate (CGC)  formalism \cite{CGC,BAL,WEIGERT},  ${\cal{N}}$ encodes all the
information about the hadronic scattering, and thus about the
non-linear and quantum effects in the hadron wave function. The
function  ${\cal{N}}$ can be obtained by solving an appropriate evolution
equation in the rapidity $y\equiv \ln (1/x)$. The main properties
of  ${\cal{N}}$ are: (a) for the interaction of a small dipole ($\rr
\ll 1/Q_{\mathrm{sat}}$),  ${\cal{N}} \ll 1$, which characterizes that
this system is weakly interacting; (b) for a large dipole
($\rr \gg 1/Q_{\mathrm{sat}}$), the system is strongly absorbed which
implies  ${\cal{N}} \approx 1$.  This property is associate to the
large density of saturated gluons in the hadron wave function.

In our analysis of diffractive heavy quark production in photon-nucleus interactions we will  consider the  phenomenological saturation model proposed in Ref. \cite{armesto} which describes the experimental data for the nuclear structure function, with the forward dipole-nucleus amplitude   parameterized as follows
\begin{eqnarray}
{\cal{N}}_A(\bar{x},\rr,\bb) = 1 - \exp \left[-\frac{1}{2}A \,T_A(\bb)\sigma_0 \, {\cal{N}}_p(\bar{x},\rr^2)\right] \,\,,
\label{enenuc}
\end{eqnarray}
where $T_A(b)$ is the nuclear profile function, which will
be obtained from a 3-parameter Fermi distribution for the nuclear
density, and  $\bar{x}= \frac{Q^2 + 4\,m_f^2}{W_{\gamma p}^2} \,.$ (For details see, e.g., Ref. \cite{vicmag_hq}). Moreover, ${\cal{N}}_p$ describes the dipole-proton
interaction. In the literature there are several phenomenological models for this quantity. Here we will consider the GBW model \cite{GBW}, which encodes the basic properties of the saturation physics and assumes that
\begin{eqnarray}
{\cal{N}}_{p} (\bar{x}, \,\rr^2) & = & 
\left[\, 1- \exp \left(-\frac{\,Q_s^2(\bar{x})\,\rr^2}{4} \right) \, \right]\,,
\label{enegbw}
\end{eqnarray}
with $ Q_s^2(\bar{x})  =  \left( \frac{x_0}{\bar{x}}
\right)^{\lambda} \,\,\mathrm{GeV}^2$ being the
saturation scale, which depends on energy and defines the onset of the
saturation phenomenon. The parameters
were obtained from a fit to the HERA data producing
$\sigma_0=23.03 \,(29.12)$ mb, $\lambda= 0.288 \, (0.277)$ and
$x_0=3.04 \cdot 10^{-4} \, (0.41 \cdot 10^{-4})$ for a 3-flavor
(4-flavor) analysis~\cite{GBW}. 
It is important to emphasize that the Eq. (\ref{enenuc}) sums up all the multiple elastic rescattering diagrams of the $q \overline{q}$ pair and is justified for large coherence length, where the transverse separation $r$ of partons in the multiparton Fock state of the photon becomes as good a conserved quantity as the angular momentum, {\it i. e.} the size of the pair $r$ becomes eigenvalue
of the scattering matrix. In our calculations of the diffractive heavy quark production in hadronic collisions we will assume that  the forward dipole-target amplitude is given by Eq. (\ref{enenuc}) in the case of a nuclear target and by Eq. (\ref{enegbw}) for a proton target.

\begin{figure}[t]
\includegraphics[scale=0.36]{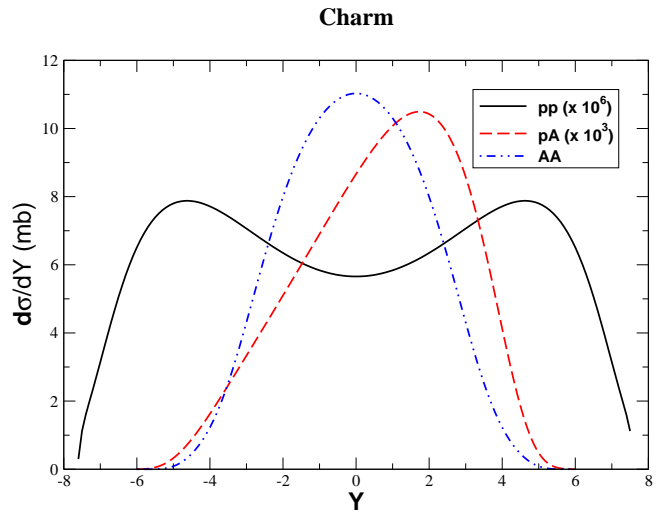}
\caption{Rapidity distribution for diffractive charm photoproduction on $pp/pA/AA$ reactions for LHC energies (see text).}
\label{fig:1}
\end{figure}

\begin{figure}[t]
\includegraphics[scale=0.36]{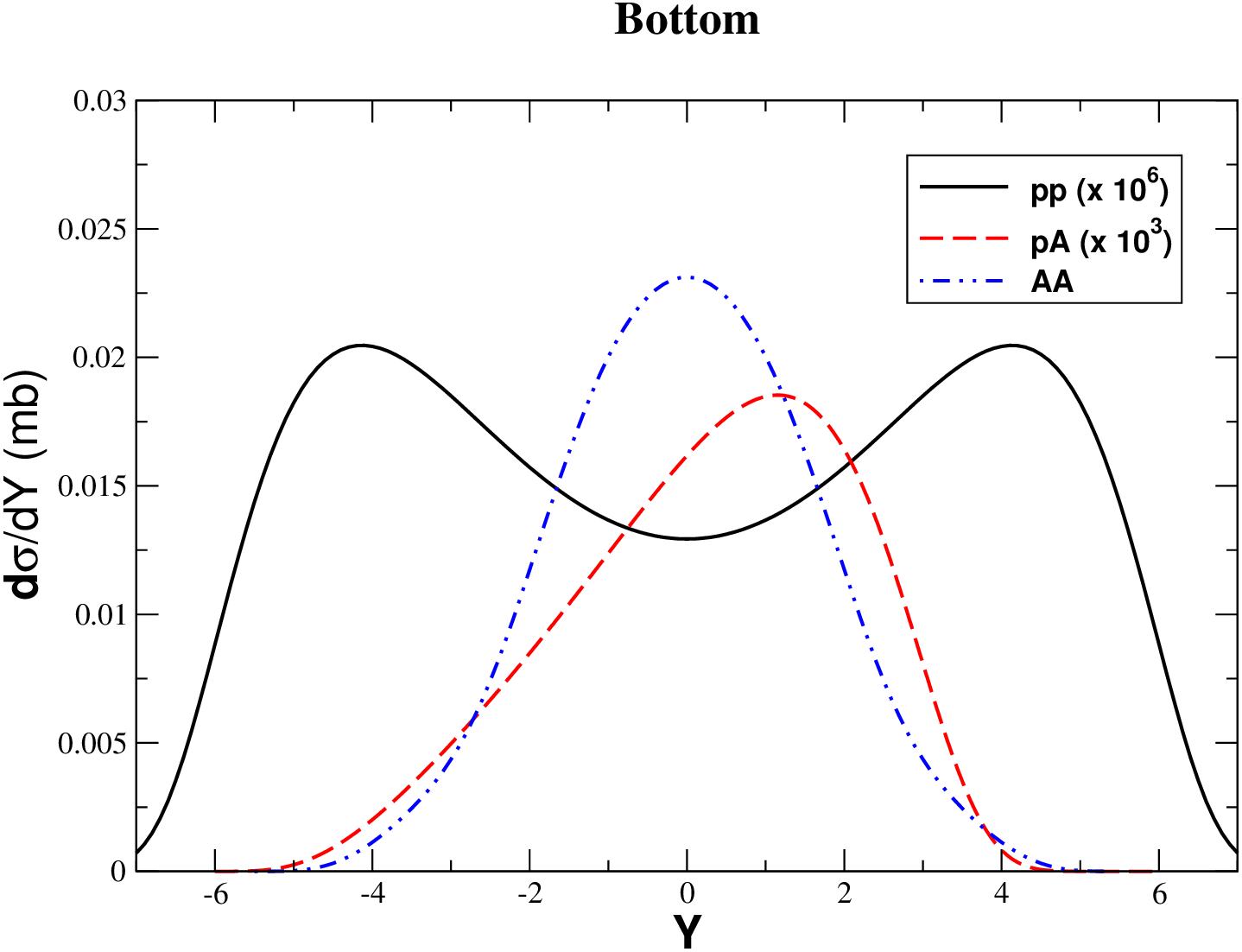}
\caption{Rapidity distribution for diffractive bottom photoproduction on $pp/pA/AA$ reactions for LHC energies (see text).}
\label{fig:2}
\end{figure}

{\it Results}. 
The distribution on rapidity $Y$ of the produced final state can be directly computed from Eq. (\ref{sigAA}), by using its  relation with the photon energy $\omega$, i.e. $Y\propto \ln \, (2 \omega/m_X)$.  Explicitly, the rapidity distribution is written down as,
\begin{eqnarray}
\frac{d\sigma \,\left[h_1 h_2 \rightarrow h_1 h_2  X \right]}{dy} = \omega \frac{dN_{\gamma} (\omega )}{d\omega }\,\sigma_{\gamma h \rightarrow X h}\left(\omega \right).
\label{dsigdy}
\end{eqnarray}
Consequently, given the photon flux, the rapidity distribution is thus a direct measure of the diffractive photoproduction cross section for a given energy.
In Figs. \ref{fig:1} and \ref{fig:2} we present our results for the diffractive heavy quark photoproduction at LHC energies.

 In Tab. \ref{tabhq} one presents the correspondent integrated cross sections. We have that the larger cross sections are obtained in the $AA$ mode, followed by $pA$ and $pp$ modes. However, the event rates should be higher in the $pp$ mode as its luminosity is several orders of magnitude larger, Namely, the  corresponding luminosities are ${\cal L}_{\mathrm{pp}} = 10^{34}$ cm$^{-2}$s$^{-1}$, ${\cal L}_{\mathrm{pPb}}=7.4\times 10^{29}$ cm$^2$s$^{-1}$ and ${\cal L}_{\mathrm{PbPb}} = 4.2 \times 10^{26}$ cm$^{-2}$s$^{-1}$. 

Let us now compare the results to processes having similar final state configuration. This analysis is important since they are competing reactions. In Ref. \cite{Heyssler}, the double diffractive (DD) production of heavy quarks has been computed (without considering rapidity gap survival correction, which diminishes the cross section). Summarizing those estimations, one has  for charm $\sigma^{DD}_{c\bar{c}} = 45 - 208 $ pb (Tevatron) and $\sigma^{DD}_{c\bar{c}}=4 - 6.56 \times 10{4}$ pb (LHC). For bottom, one has $\sigma^{DD}_{b\bar{b}} = 17 - 78$ pb (Tevatron) and $\sigma^{DD}_{b\bar{b}} =0.5 - 1.5 \times 10^{4}$ pb (LHC). Our result for the $pp$ mode are at least one order of magnitude larger. Other process with similar configuration  is the double photon process in the $AA$ mode. In Ref. \cite{per2}, we obtain the following values for coherent PbPb collision at LHC energies: for charm, $\sigma_{c\bar{c}}^{\gamma\gamma}= 1.8$ mb and for bottom $\sigma_{c\bar{c}}^{\gamma\gamma}=2$ $\mu$b. Our results are higher by a factor 30 for charm and a factor 5 for bottom. 

{\it Summary}. The QCD dynamics at high energies is of utmost importance for building a realistic description of $pp/pA/AA$ collisions at LHC. In this limit QCD evolution leads to a system with high gluon density. 
In this letter we have studied the diffractive photoproduction of heavy quarks, which provide a feasible and clear measurement of the underlying QCD dynamics at high energies.  The advantages of this process are the clear final state (rapidity gaps and low momenta particles) and no competing effect of dense nuclear environment if compared with hadroproduction. However, as the present analysis is predominantly phenomenological several points deserve more detailed studies. For instance, the model dependence as well as estimative of background processes and the analysis of the experimental separation has to be further addressed.

It is important to emphasize that the same reaction, $h_1 h_2 \rightarrow h_1 h_1 X$, also occurs via fusion of two Pomerons, the so-called central diffraction processes. However, the transverse momenta of the scattered hadrons are predicted to be much larger than in two-photon interactions, which implies that the separation between these two processes is feasible. In the case of diffractive photoproduction we expect an asymmetric distribution of the scattered hadrons, since photon and pomeron exchange are present in the process. Moreover, as almost all of the photoproduced heavy quarks, similarly to the vector mesons, should have small transverse momenta, it is possible to introduce a selection criterion to separate the diffractive photoproduction processes.

\vspace{-0.6cm}

\begin{acknowledgments}
This work was partially financed by the Brazilian funding agencies CNPq and FAPERGS.
\end{acknowledgments}

\end{document}